\documentclass[sigconf]{acmart} 

\copyrightyear{2026}
\acmYear{2026}
\setcopyright{cc}
\setcctype{by}
\acmConference[HT '26]{37th ACM Conference on Hypertext}{September 14--18, 2026}{London, United Kingdom}
\acmBooktitle{37th ACM Conference on Hypertext (HT '26), September 14--18, 2026, London, United Kingdom}
\acmDOI{10.1145/3800935.3830891}
\acmISBN{979-8-4007-2564-7/2026/09}

\begin{document}

\title{Creative Reading: Scaffolding Reading for Transformation}



\author{Sophia Liu}
\orcid{0009-0008-7746-0749}
\affiliation{%
  \institution{University of California, Berkeley}
  \city{Berkeley}
  \state{California}
  \country{USA}
}
\email{sophiawliu@berkeley.edu}

\author{Sarah Abowitz}
\orcid{0009-0000-1034-303X}
\affiliation{%
  \institution{Tufts University}
  \city{Medford}
  \state{Massachusetts}
  \country{USA}
}
\email{sarah.abowitz@tufts.edu}


\author{Yijun Liu}
\orcid{0009-0008-6601-9237}
\affiliation{%
  \institution{University of Illinois Urbana-Champaign}
  \city{Urbana}
  \state{Illinois}
  \country{USA}
}
\email{yijun6@illinois.edu}

\author{Sarah Sterman}
\orcid{0000-0002-9282-559X}
\affiliation{%
  \institution{University of Illinois Urbana-Champaign}
  \city{Urbana}
  \state{Illinois}
  \country{USA}
}
\email{ssterman@illinois.edu}

\author{Shm Garanganao Almeda}
\orcid{0000-0001-7660-313X}
\affiliation{%
  \institution{University of California, Berkeley}
  \city{Berkeley}
  \state{California}
  \country{USA}
}
\email{shm.almeda@berkeley.edu}

\author{Max Kreminski}
\orcid{0009-0002-6268-4033}
\affiliation{%
  \institution{Cornell Tech}
  \city{New York}
  \state{New York}
  \country{USA}
}
\email{mkremins@cornell.edu}


\begin{abstract}
Reading augmentation systems increasingly help readers process text at scale. While these tools address real constraints of time and cognitive load, they often implicitly frame reading as information transmission, or ``reading to discard,'' delegating interpretation and effort to the machine. Yet this delegation changes the outcome of reading. For example, in scholarly reading, deciding what a research text implies and why it matters is central to the work of scholarly production. We propose \textit{creative reading} as an alternative goal: reading augmentation that supports readers in creating both readings and themselves as readers. By putting literary and narrative theories into conversation with scholarly sensemaking and creativity support, we present a provocation-oriented design space for valuing the process of reading as a way of preserving a plurality of readings and transforming readers over time.
\end{abstract}

\ccsdesc[500]{Human-centered computing~Interaction paradigms}


\keywords{augmenting reading, scholarly sensemaking, creative reading, hypertext reading, creativity support}

\begin{teaserfigure}
  \includegraphics[width=\linewidth]{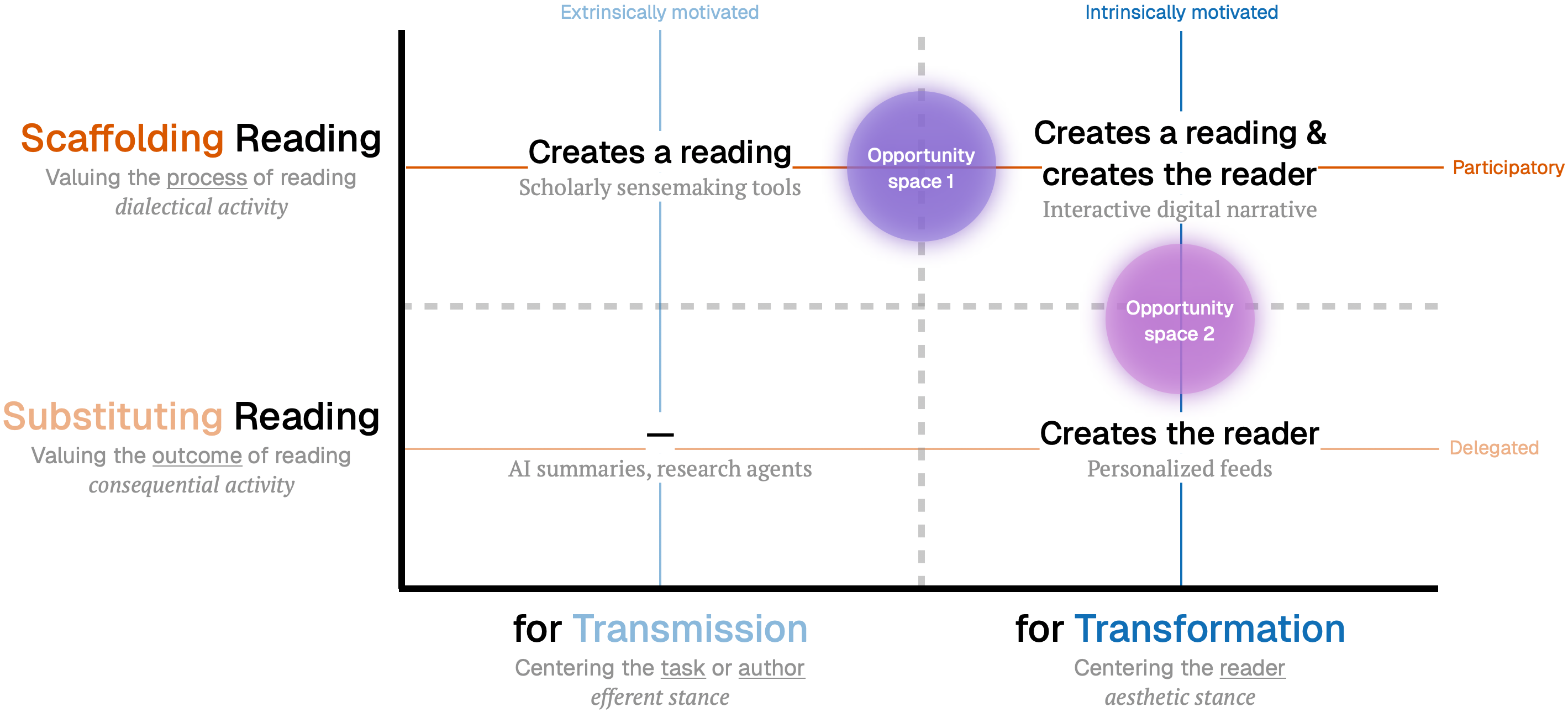}
  \caption{A provocation-oriented design space for creative reading, mapping systems along two axes: reading for \textit{transmission} to reading for \textit{transformation}, and \textit{substituting} to \textit{scaffolding} the reading process. While many existing reading augmentation systems for scholarly sensemaking center efficient transmission, we look toward two opportunity spaces that increase reader agency: (1) transformative scholarly scaffolding and (2) more participatory algorithmic curation.}
  \Description{This figure presents a two-dimensional design space for reading augmentation systems. On a white background, there is a black horizontal axis at the bottom of the figure, and a black vertical axis on the left of the figure. Each axis is bisected by a grey dotted line that splits the space into four quadrants. The horizontal axis begins from one extreme of reading for transmission on the left, labeled in a lighter blue, and proceeds toward reading for transformation on the right, labeled in more of a deeper azure. Under the transmission label, grey text reads "Centering the task or author (efferent stance)". Under the transformation label, grey text reads "Centering the reader (aesthetic stance)". Lines in each of these corresponding blues reach up from the horizontal axis to the top of the figure, bisecting the space on that axis between an end and the dotted line in the middle. The light blue line above "for transmission" terminates in light blue text reading "Extrinsically motivated" and the deeper azure line above "for transformation" terminates in deeper azure text reading "Intrinsically motivated." The vertical axis begins from one extreme of substituting reading, labeled in a pale red at the bottom, rising toward scaffolding reading, labeled in a brighter red. Under the substituting label, grey text reads "Valuing the outcome of reading (consequential activity)." Under the scaffolding label, grey text reads "Valuing the process of reading (dialectical activity)." Lines in each of these corresponding reds reach across from the horizontal axis to the right of the figure, bisecting the space on that axis between an end and the dotted line in the middle. The pale red line next to "substituting reading" terminates in pale red text reading "Delegated" and the brighter red line next to "scaffolding reading" terminates in brighter red text reading "Participatory." Each quadrant formed by the dotted lines are also labeled with black text. In the upper-left quadrant, the label reads, "Creates a reading: scholarly sensemaking tools." In the lower-left quadrant, the label reads "AI summaries, research agents," with the lack of a title for this label suggesting they do not meaningfully create a reading or the reader. In the upper-right quadrant, the label reads "Creates a reading \& creates the reader: Interactive Digital Narrative (IDN)", which is generally associated with augmented reading that is participatory, intrinsically motivated, and oriented toward transformation. In the lower-right quadrant, the label reads "Creates the reader: personalized feeds", which is generally associated with reader transformation through delegated systems rather than participatory reading. Two translucent circular ``opportunity spaces'' are highlighted. Opportunity space 1 is a blue-violet circle that sits where the brighter red line crosses the vertical dotted line, suggesting a possible design direction for systems that better scaffold reading while moving from transmission toward transformation. Opportunity space 2 is a red-violet circle that sits where the deeper azure line and the horizontal dotted line intersect, suggesting another design direction around transformative systems that shape the reader, but with attention to whether that transformation is participatory or delegated.}
  \label{fig:teaser}
\end{teaserfigure}

\maketitle
\section{Introduction}

Reading can be an active, creative, and transformative encounter with a text. We read for pleasure; seek out difficult texts in order to expand our own readerly capacities; read slowly to savor the feeling of a text; revisit texts to see how their meaning has changed for us over time; juxtapose texts to see what they draw out of one another; read to engage with the author; scrutinize texts and develop opinions; and, in general, read \emph{creatively}, forging our own paths and assembling our own interpretations within and across the texts we select. In these contexts, readers make meaning and are changed by the process. \textit{Augmentation} can support the readerly capacities readers develop in their engagements with texts.

However, recent work on augmenting reading largely focuses on making reading more \emph{efficient}, especially in knowledge-intensive contexts where people must rapidly make sense of research papers and other complex documents~\cite{SemanticReaderProject}. These systems often aim to help readers skim, summarize, and synthesize large volumes of information in order to reduce cognitive effort and accelerate document- and corpus-level sensemaking at scale~\cite{AbstractExplorer, SemanticReaderProject, Fok2023Scim}. At one extreme, the form of reading implicitly envisioned by these systems may resemble what we call ``reading to discard'': extracting the essential, actionable informational \emph{core} of a document as quickly and efficiently as possible, so that the rest can be swiftly tossed aside---much like we often process inbound mail. This motivates us to ask what scaffolding approaches might resist ``reading to discard,'' what alternative forms of reading we might scaffold, and what alternative \emph{views} and \emph{values} of reading might guide the conversation.

Hypertext and interactive digital narrative have long explored active forms of reading~\cite{Murray1997, Bernstein2009HypertextNarrative}. In these traditions, meaning is developed by and for the reader through movement, choice, and repeated engagement across linked and procedural textual spaces~\cite{Aarseth1997, Twine2022, Koenitz2015IDN}. Work framing hypertext as a method of inquiry and augmentation further shows how reading can be a way of experiencing and thinking through relations rather than simply retrieving content~\cite{Atzenbeck2019HypertextAsMethod}. Our move is to bring these ideas into scholarly reading augmentation, where readers of research texts often have practical goals, while interpretation and reader development remain central to the work of reading and sensemaking.

We draw on these ideas to reconsider reading augmentation beyond paradigms of transmission and ``reading to discard,'' in which texts are treated primarily as sources of extractable information. Rather than merely carrying knowledge away from texts, readers may instead participate in \emph{creative} and \emph{transformative} forms of reading through which they simultaneously create both ``readings'' (Sec.~\ref{sec:creating_a_reading}) and themselves as readers (Sec.~\ref{sec:creating_a_reader}). We develop this view by bringing literary theory and interactive digital narrative into conversation with scholarly sensemaking and creativity support, showing why augmentation should support the creation of readings and the development of readers. We then present a provocation-oriented design space (Fig.~\ref{fig:teaser}) that clarifies what existing systems value and surfaces opportunities for reading augmentation that center reader process and transformation.


\section{Creating a Reading}
\label{sec:creating_a_reading}
Longstanding theories of interpretation in literary criticism and hermeneutics suggest that reading can be creative when the encounter between reader and text produces something that neither fully contains in advance~\cite{Barthes1974SZ, ecoReader}. Drawing on Rosenblatt's transactional theory of reading~\cite{rosenblatt1938literature}, we call the product of this transaction ``a reading,'' the interpretation or artifact created during the activity of reading. Thus, we present reading as both a creative process and its outcome. This artifact is distinct from a later work that uses information gained from the reading. It may remain internal or nebulous as associations and interpretations, or become partially externalized through traces such as manual marginalia or computationally logged paths reminiscent of read-wear~\cite{Hill1992EditWearAndReadWear, Bernstein2025BackToThe}. A reading may take the form of a personal interpretation of a novel, a chain of associations across linked Wikipedia articles, a set of annotations accumulated while studying a paper, or an evolving synthesis constructed across a research corpus.

\subsection{Reading as Co-Creative Activity}
Barthes helps clarify why reading should be understood as creative. Through his distinction between ``readerly'' and ``writerly'' texts~\cite{Barthes1974SZ}, he contrasts texts that position the reader as a consumer of fixed meaning with texts that invite the reader into the production of meaning. In ``The Death of the Author,'' he declares this as the ``birth of the reader''~\cite{barthes1977death}. Thus, if the meaning of a text is co-created with the reader, then reading augmentation should support readers in creating what we call a reading.

Reader-response theorists similarly emphasize the active and creative role of the reader. Iser describes literary works as existing between an ``artistic'' pole, the text produced by the author, and an ``aesthetic'' pole, the realization of the work by the reader~\cite{iser1974implied}. The work sits ``half-way between the two'' as the process of reading engages the static text's ``inherently dynamic character'' through the reader's unique and evolving interpretation. Texts alone contain interpretive ``gaps'' that readers fill in through their imagination and synthesis with prior readings. He calls this convergence the work's ``virtual dimension,'' or ``the coming together of text and imagination'' in a new \textit{co-creation} between author and reader.

Hypertext operationalized many of the participatory qualities Barthes envisioned in his ideal text, enabling a plurality of interpretations from reader-led interaction with dynamic textual systems. In interactive digital narrative (IDN) systems, readers shape their interaction within a narrative through their navigation of the narrative system. Murray describes this authorial capacity in IDN as \emph{agency}~\cite{Murray1997}. She makes the critical distinction between agency and authorship, arguing that ``the interactor is not the author of the digital narrative, although the interactor can experience one of the most exciting aspects of artistic creation---the thrill of exerting power over enticing and plastic materials. This is not authorship but agency.'' From this perspective, readers of IDN may not author the narrative system, but they still improvise as ``interactors'' within its pre-authored, material-like possibility space.

\subsection{Two Reading Stances}
Rosenblatt gives us a way to carry these ideas of reader agency beyond literary and explicitly interactive texts. Although reader-response theory and IDN scholarship are primarily concerned with literary texts, Rosenblatt argues that the way a reader approaches a text is not determined solely by the type of text~\cite{rosenblatt1938literature}. Readers may adopt different reading stances toward the same text, attending to and valuing different aspects of the reading experience. When reading from an \emph{efferent}\footnote{Rosenblatt uses \emph{efferent} from the Latin \textit{efferre}, ``to carry away.''} stance, readers focus on the information they will carry away from the text, such as facts, instructions, or claims. In contrast, an \emph{aesthetic} stance emphasizes the lived-through experience of reading, including the feelings, images, and associations produced during the reader's engagement with the text, similar to Iser's framing of the term. In the next section, we discuss how this distinction matters for reading augmentation, and how system design can encourage different stances toward the same text.

\subsection{The Forking Path}
This distinction parallels a broader divergence within the evolution of reading augmentation systems. Early hypertext visionaries foresaw the liberatory potential of non-linear information systems in informational, educational, and creative contexts~\cite{Nelson1987LiteraryMachines, Bush1945As, Stephens}. As digital media became associated with attention capture, informational simplification~\cite{isGoogleMakingUsStupid}, and cultural homogenization~\cite{algorithmicHomogeneity}, Kolb et al. reimagined hypertext's potential as ``resistance'' to these forces, calling for texts that demand from the reader time, effort, and attention beyond the immediate horizon of an individual text~\cite{hypertextResistance}.

Across this history, reading augmentation has increasingly split between efficient information transmission and open-ended, interpretive exploration. We argue that contemporary systems for \textit{scholarly sensemaking} (i.e., making sense of research papers and ideas in order to organize and synthesize scholarly knowledge~\cite{Zhu2024Patterns, Zhu2023ScholarlyKnowledgeSynthesis}) have optimized for the \emph{efferent} reading stance, often treating the goal of reading as efficiently extracting knowledge from large and complex volumes of text~\cite{Fok2023Scim, AbstractExplorer, Head2021Augmenting}, or ``reading to discard.'' Conversely, hypertext fiction and IDN systems have more often emphasized the \emph{aesthetic} process of open-ended, interpretive exploration, where readers experience agency in determining the value they take away from the text~\cite{Koenitz2015IDN, Murray1997, Twine2022, Bernstein2009HypertextNarrative}.


Recent hypertext scholarship has revisited the historical divergence between hypertext as a medium for productivity and as a medium for entertainment~\cite{Anderson2023Seven}. Our work argues that these paths should not remain separate. Scholarly reading often has \emph{efferent} goals because readers need to carry knowledge forward into their own work. Yet even when the goal is efferent, the process is often aesthetic. To synthesize a field or develop a hypothesis, readers must move across and revisit texts, notice relations among them, and gradually construct their own position from their unique readings. Treating hypertext as a method of inquiry~\cite{Atzenbeck2019HypertextAsMethod}, we apply these literary theories to scholarly contexts and argue that reading augmentation systems should support an aesthetic reading process even to achieve efferent reading goals.

Recent HCI work on scholarly sensemaking describes reading and synthesis as an iterative process across interconnected documents and notes~\cite{Zhu2024Patterns, Zhu2023ScholarlyKnowledgeSynthesis}. Work in information search similarly suggests that users often prefer indirect, incremental, and contextualized navigation moves toward an answer over direct and instant but often de-contextualized ones~\cite{Teevan2004Orienteering, ODay1993Orienteering}. Even efferent, goal-oriented reading involves forms of open-ended exploration and interpretation that resemble the aesthetic reading experience, seemingly reserved for literary texts with more room for subjective, reader-centric experience. 

We posit that augmentation of ``objective'' scholarly works should therefore consider reading ideals that go beyond information \textit{transmission} and toward reader \textit{transformation}. In an efferent model, reading carries information from the text to the reader, while in an aesthetic model, reading affects the reader through their interaction with the text. Barthes' ``birth of the reader'' elevates the reader to an active participant in the meaning of a text~\cite{barthes1977death}. We extend this birth from the production of meaning into a process of becoming. Reading creates readings, and in doing so, creates readers.

\section{Creating the Reader}
\label{sec:creating_a_reader}
\subsection{Reading for Transformation}
The previous section described the creation of a reading; this section turns to the reader created through that process. A reading is shaped by a reader's interaction with a text; that interaction also shapes the reader. Because the aesthetic experience of reading is lived through by the reader, it cannot be completely delegated to another reader or system~\cite{rosenblatt1938literature}. This makes reading augmentation a question of supporting the reader's capacities, rather than outsourcing the process of reading, and marks a return to Engelbart's broader vision of augmentation: rather than ``isolated clever tricks that help in particular situations,'' augmentation can support ``a way of life in an integrated domain,'' where complex problems may unfold ``for twenty minutes or twenty years''~\cite{engelbart1962augmenting}. Engelbart emphasizes that augmentation should change a person's ongoing way of working across time and contexts. Reading augmentation should therefore move beyond faster information transmission for specific tasks toward the development of deeper readerly capacities over time.

We call our alternative vision \textit{reading for transformation}: reading augmentation that treats the reader's transformation as its central design concern. Any augmentation of reading changes the reader in the moment of reading. In an efferent stance, that change is often bounded to the completion of a particular reading goal. Reading for transformation instead asks how systems might encourage an aesthetic stance toward reading, scaffolding changes that exceed the immediate task by shaping how readers read and think, and consequently, the questions, tastes, and sensibilities they bring to future readings and work.

Glăveanu and Beghetto's non-standard definition of creativity through ``creative experience'' supports our view of reading as a creative activity. They describe creative experience as a novel person--world encounter grounded in meaningful action and interaction, marked by open-endedness, nonlinearity, pluri-perspectives, and future-orientation~\cite{Glaveanu2021CreativeExperience}. These qualities are central to our conception of reading for transformation. Readers approach texts without knowing in advance what they will carry away, move through them in nonlinear and dynamic trajectories, bring multiple perspectives into relation with one another, and connect present reading to future thought and work. Yet future-orientation goes beyond what the reader will later produce. It also concerns the future reader who comes into being through the act of reading. Readers select and approach texts in relation to who they might become. Reading, then, is a person--world encounter through which the reader's orientation toward texts, ideas, and themselves changes, and a deeply creative and transformative experience.

\subsection{Orienteering as Reader Agency}
We may also consider this change in \textit{orientation} more directly, via Murray's account of navigation as a source of reader agency. Murray argues that navigating digital environments can be pleasurable, as readers orient by landmarks and build mental maps~\cite{Murray1997}. She compares this to the sport of orienteering, where participants navigate complex terrain by following clues at intermediate steps. She explains that IDN systems offer this pleasure in the two configurations of the solvable maze, where movement is directed toward a known destination, and the tangled rhizome, where navigation remains open-ended. Murray also distinguishes instrumental navigation (toward necessary destinations like a dentist's office or airport gate) from pleasurable navigation done ``for sport.''

However, information search research suggests that this distinction is not so clear-cut. In a study of orienteering behavior in directed search, a participant looking for a professor's office number does not ``teleport'' directly to the answer, even when that option exists. Instead, they move through the departmental homepage, a people-finding directory, and a list of faculty, in a process the authors also call orienteering~\cite{Teevan2004Orienteering}. The route matters even when direct answers are available; it lets the user see the answer in context and form a useful path they can retrace mentally or physically. Orienteering illustrates why readers may need to move through the process of reading themselves, even when delegation tools are available. Transmission can deliver an answer more quickly, but transformation depends on navigating the space that gives it meaning, exemplifying hypertext as method~\cite{Atzenbeck2019HypertextAsMethod}.

Orienteering is pleasurable, like many sports, because effort is integral to the activity. Reaching the destination fastest may be the goal of the game, but much of the pleasure is derived from the process of orienteering toward it. The difficulty of this navigational work is a part of its transformative power. Murray describes how moving through a complex narrative space can feel like an enactment of courage and perseverance, experienced by the reader as much as by a story's protagonist~\cite{Murray1997}. Work on creative struggle in creative writing similarly frames effortful engagement as a condition for critical reflection, self-dialogue, and deeper self-understanding~\cite{CreativeStruggle}. In reading augmentation, productive friction has been proposed to strategically preserve or introduce effort that supports deeper comprehension and engagement~\cite{ProductiveFriction}. 

Zhang's definition of dialectical activity clarifies why this effort matters beyond comprehension alone. In a consequentialist view, an activity is valuable because it produces a desired outcome, such as a problem solved or a measurable result achieved. Dialectical activities, by contrast, are valuable because their good can only be discovered from within the activity, as in parenting, art-making, friendship, or research~\cite{Zhang2024}. Scholarly sensemaking is similarly dialectical: researchers develop skill, perseverance, and perspective---and become better researchers---by orienteering through an open-ended field of work themselves.

\subsection{Toward Process Aesthetics for Reading}
To be transformed by reading, readers must participate in the effort, uncertainty, and pleasure through which they create themselves over time. This emphasis on reading as a process-valued activity connects reading augmentation to \textit{process aesthetics} in creativity support research: the experiential qualities of using a creativity support tool, distinct from the qualities of the artifact it helps produce~\cite{Kreminski2021ReflectiveC}. For creative reading, we point to two particularly useful process aesthetics: flow and reflection.

Flow helps readers stay with texts by making effort feel pleasurable, expressive, and intrinsically motivating, as in Compton's description of casual creators~\cite{Compton2019}---often through agency and responsive interaction~\cite{Nakamura2002ConceptFlow}. Reflection, as in Kreminski \& Mateas' reflective creators~\cite{Kreminski2021ReflectiveC}, helps that effort become transformative by facilitating more contemplative forms of reading. Interventions designed to emphasize flow and reflection in the reading process could help readers move through long or difficult texts, and be changed by them, even when delegation is available.

We employ Nakakoji's analogies of creativity support tools as \textit{running shoes}, \textit{dumbbells}, or \textit{skis} to help translate these desired process aesthetics into goals for reading augmentation system design~\cite{nakakoji2006meanings}. Running-shoe systems can reduce the frictions that impede aesthetic reading, helping readers create readings more effectively, and more often. Dumbbell systems develop and sustain readerly capacities through skill-strengthening and reflection, helping create the reader. Ski systems enable new forms of reading that would be difficult to otherwise imagine. We call this broader ambition \textit{creative reading}: reading augmentation that supports readers in creating \textit{readings}, and in creating themselves as \textit{readers}.

\section{A Design Space for Creative Reading}

Creative reading gives us a lens for organizing a provocation-oriented design space of existing and speculative reading augmentation systems. In this paper, reading refers to the situated activity of encountering, navigating, interpreting, and relating texts, both within and across sources. We focus on scholarly reading and sensemaking because it combines clear efferent goals with deeply interpretive, open-ended activity. Readers must carry knowledge forward in their own work, but they do so by building perspectives over long horizons. Following Vitali's call to ask what conception of reading is assumed, what is optimized and for whom, and what may be diminished through augmentation~\cite{vitali2026whatareweDOING}, we map systems by the values their interfaces and evaluations foreground. All reading may transform readers in some way; our distinction is whether that transformation remains incidental or bounded to task success, or whether the system is designed around open-ended change in the reader. Like prior HCI design spaces~\cite{Lee2024ADesignSpace, kim2017mosaic} and provocation-oriented affinity maps~\cite{So2026Beyond}, this space aims to identify dominant assumptions and surface underexplored opportunities.

Figure~\ref{fig:teaser} maps reading augmentation systems along two axes. The horizontal axis asks what reading is oriented toward. Systems that support reading for \textit{transmission} center the task or author and promote an efferent reading stance, where value lies in what the reader carries away. Systems that support reading for \textit{transformation} center the reader and promote an aesthetic reading stance, where value lies in the reader's experience and changing orientation during and beyond the reading. The vertical axis differentiates minimalist or offloaded reading from maximalist or participation-heavy reading. In \textit{substituting} reading, systems value the outcome of reading and perform reading work on behalf of readers. In \textit{scaffolding} reading, systems value the process of reading and support the reader's own activity. We use these axes to reveal four orientations for reading augmentation and two opportunity spaces for designing creative reading systems in the transitions between them.

\subsection{Four Orientations in Reading Augmentation}
\subsubsection{Substituting Reading for Transmission}

AI summaries~\cite{xu2025aisummariesonlinesearch}, document question-answering systems~\cite{Hou2026DocQA}, and research agents~\cite{huang2025deepresearchagentssystematic} perform reading work for readers so they can quickly carry away the task-relevant content or primary message of a text. When reading is valued primarily for this outcome, delegation is useful for tasks where direct engagement with the text is unnecessary, such as checking a fact or finding a definition. Yet the boundary is often blurry~\cite{vitali2026whatareweDOING}. The same system can substitute for reading when process and interpretation matter, such as when a student writes an essay on a classic novel from summaries or a researcher cites a paper they have only queried through an agent. The system may produce an outcome without creating a reading or developing the reader, because the reader receives an interpretation without experiencing the process that would make it their own.

\subsubsection{Scaffolding Reading for Transmission}

Scaffolding for transmission keeps readers in the work of reading while supporting bounded, efferent goals. This orientation can resemble ``reading to discard,'' where readers working under constraints of time and attention use summaries and other cognitive scaffolds to decide what deserves closer engagement. Like running shoes, these systems support familiar reading tasks with relatively clear criteria for success, such as skimming and comparing~\cite{nakakoji2006meanings, vitali2026whatareweDOING}. Their practical value lies in reducing frictions that impede reading. Good scaffolds reduce cognitive load while leaving the reader responsible for creating a reading.

Existing scholarly reading systems primarily sit in this orientation. \textsc{Scim} guides attention to predefined content types in academic papers~\cite{Fok2023Scim}. \textsc{ScholarPhi} provides just-in-time definitions so readers can stay within the flow of a paper~\cite{Head2021Augmenting}. \textsc{Relatedly} organizes related work sections so readers can more efficiently and effectively explore the scope of a literature when conducting literature reviews~\cite{Palani2023Relatedly}. As scaffolds, these systems are participatory, yet they remain transmission-oriented because their interfaces and evaluations foreground better scholarly products. \textsc{AbstractExplorer} sits near the edge of this orientation because it supports comparative close reading as a dialectical activity~\cite{AbstractExplorer, Zhang2024}. It preserves author-written sentences and surfaces variation across them for readers to identify themselves, but its evaluation still foregrounds bounded tasks such as corpus familiarization and filtering toward personally relevant papers. It points toward transformation but is evaluated mainly through transmission. Scaffolding for transmission helps readers make better readings of texts; scaffolding for transformation asks how systems might make room for the time and agency through which those readings change the reader.

\subsubsection{Scaffolding Reading for Transformation}

Scaffolding for transformation takes the motivation behind dialectical reading and runs with it. It carries the impulse behind \textsc{AbstractExplorer} one step further, from repeated engagement for better comprehension of a corpus, toward repeated engagement for reader transformation. In designing for transformation, the reading process is valued from within, the essence of dialectical activity~\cite{Zhang2024}, so augmentation must work across longer, more indefinite horizons than a single outcome. It centers the reader's development across texts and projects as repeated engagement shapes how readers read. Like dumbbells, these systems can strengthen readerly capacities. Like skis, they can enable new ways of moving through texts~\cite{nakakoji2006meanings} for greater reader immersion, agency, and transformation~\cite{Murray1997}. They are more faithful to Engelbart's definition of augmentation, treating it as more than a way of working and closer to a way of being.

Though primarily existing in narrative and non-scientific contexts, IDN offers the clearest model for this orientation because the reading's value lies in the aesthetic experience of moving through a textual possibility space. To recall Murray, readers experience agency by improvising within the narrative system~\cite{Murray1997}. \textsc{Twine} and other nonlinear story systems encode that agency through branching paths that let readers experience the effects of their choices~\cite{Carlon2022Twine}. Reading IDN can be pleasurable and self-motivating~\cite{Murray1997, Compton2019}, while (re)reading can also induce reflection on paths taken and not taken~\cite{Bernstein2009HypertextNarrative, Kreminski2021ReflectiveC}. The lesson for supporting scholarly sensemaking is to bring these process aesthetics into the expository context of research literature. Reading augmentation can scaffold orienteering through a field of knowledge, support readers in creating paths across ideas~\cite{Zhu2024Patterns}, and help them develop an orientation they carry beyond their immediate reading goals.

Some scholarly sensemaking systems already gesture toward this orientation. \textsc{LiquidText} supports \textit{active reading} by making documents spatially manipulable, letting readers arrange and annotate them for comparison and synthesis~\cite{Tashman2011Liquid}. Its flexible reading paradigm recalls spatial hypertext's use of layout to express meaning~\cite{Anderson2025Whither}, honoring hypertext's original ambition to move beyond paper constraints~\cite{Bush1945As, Nelson1987LiteraryMachines} and toward new ways of participating in and thinking with texts~\cite{Atzenbeck2019HypertextAsMethod}. Hypertext notebooks are not reading augmentation systems in the narrow sense of augmenting text consumption, but they matter for reader transformation because they extend reading into the long-horizon work of interpretation across texts and projects. By preserving traces of prior readings, they help readers reflect on and reuse ideas gleaned from them, developing the orientations they bring to future work~\cite{Zhu2024Patterns}. These systems are more ski-like because they enable new movements through texts---both literally and figuratively. \textsc{Tompkinsia}, by contrast, is more dumbbell-like, helping language learners and scholars alike develop language skills and critical judgment through linked source text, commentary, and data-driven analyses, recalling commentary's older role in scaffolding the aesthetic experience of difficult historical texts~\cite{abowitz2026tompkinsia}.

Still, these systems only partially realize the orientation that IDN encourages. They support and preserve the creation of readings, foster navigational and spatial thinking, and develop readerly skills, but they do not fully design scholarly reading around the experiential qualities of process aesthetics. There is an opportunity to support reader transformation more directly, helping readers develop the courage, tastes, and capacities through which they continually create themselves as readers. This suggests Opportunity Space 1 in Figure~\ref{fig:teaser}: designing for \textit{transformative scholarly scaffolding}, or supporting the experiential qualities of reading central to developing readerly capacities over time.

\subsubsection{Substituting Reading for Transformation}

Substituting reading for transformation is a prolific area of this design space, and one that we urge designers to consider cautiously. This orientation is embedding in systems that design for the process aesthetic of flow without reflection; when systems shape what readers think and value while neglecting, unintentionally or otherwise, their agency and participation in that process of change. AI search summaries can bias users' attitudes and policy support toward the stance presented in the summary~\cite{xu2025aisummariesonlinesearch}; link recommendation algorithms can increase ideological clustering in echo chambers and political polarization by guiding users' social paths~\cite{Santos2021Algorithms}; and sycophantic AI advice can affirm users in ways that reduce critical and self-correcting behavior~\cite{Cheng2026Sychophantic}. This final quadrant shows the risk of delegating the creative process of reader transformation, where systems change readers through hedonic, intrinsically motivating~\cite{Bennett2024Intrinsic} experiences without the agency and reflection through which readers create themselves. While this paper primarily focuses on Opportunity Space 1, this risk also suggests Opportunity Space 2: addressing \textit{algorithmic curation as reader creation} by moving substitutive systems toward greater reader participation, as explored in prior work~\cite{Liu2025Agency, ElMalki2026Bonsai}.

\section{Conclusion}
\begin{quote}
\textit{If educators wish to foster students’ critical reading and thinking abilities, then certainly students need to be able to develop, trust, and give voice to their own aesthetic experiences with literature. However, if students are taught that the goal of reading, even for literary texts, is to extract a correct, public meaning (usually one established by the teachers’ guide), they will adopt efferent stances for reading that promote factual rather than thoughtful comprehension, inhibit critical-thinking skills, and limit preparation for enfranchisement in a democracy.}
\end{quote}
\begin{flushright}---Angela S. Raines~\cite{Raines2005Rosenblatt}\end{flushright}

Reading augmentation, even in scholarly contexts, should not be reduced to more efficient information transmission. This paper began from a divergence between systems that help readers carry information away from texts and traditions in reader-response theory, hypertext, and IDN that understand reading as an aesthetic and participatory process. We argue that readers create \textit{readings} by authoring their own paths and interpretations, and in creating readings, also create \textit{themselves}, developing the orientations and capacities they bring to future texts.

We contribute a legible and applicable articulation of this perspective as a provocation-oriented design space, identifying \textbf{substitution} and \textbf{scaffolding}, \textbf{transmission} and \textbf{transformation} as useful orientations for different reading contexts. We seek opportunities that go beyond a superficial rejection of efficiency, while resisting a monoculture that treats all goal-oriented reading as transmission-oriented or ``reading to discard.'' Our conception of creative reading points toward augmentation systems that help readers make readings their own, and that support the longer, uncertain, pleasurable, and reflective processes through which those readings can transform the reader.

This paper has focused on the individual reader, but creative reading has profoundly collective stakes~\cite{cosmikVision}. Returning to the reader-response literary school of Rosenblatt and Iser, Fish reminds us that interpretation is also shaped by the ``interpretive communities'' we are part of~\cite{fish1980text}. Engelbart's original vision of augmentation was also famously collective, chiefly concerned with how people think and work together on evolving problems over time~\cite{engelbart1962augmenting}. Hypertext is a powerful medium and method here because it can produce and preserve a plurality of readings through visible links and traces~\cite{UISTful}, rather than collapsing interpretation into a single takeaway via AI summaries or siloed interpretive communities. Raines' warning similarly suggests that teaching readers to extract a single public meaning can inhibit critical thought and democratic participation, even and especially beyond the classroom~\cite{Raines2005Rosenblatt}. In that same vein, reading augmentation systems should help readers ``develop, trust, and give voice to their own aesthetic experiences'' while bringing that plurality into relation with others.

\begin{acks}
This work emerged from a discussion group at the Science and Technology for Augmenting Reading (STAR) workshop~\cite{STAR2026} at CHI 2026. Thanks to the STAR organizers for making this discussion group possible, and to Eric Rawn and Borano Llana for their contributions during the workshop.
\end{acks}

\bibliographystyle{ACM-Reference-Format}
\bibliography{bibliography}


\end{document}